\documentclass[12pt]{article}

\usepackage{epsf}
\usepackage[dvips]{graphicx}

\begin{document}

\begin{center}
{\bfseries  TENSOR $A_{yy}$ AND VECTOR $A_y$ ANALYZING POWERS OF
THE $(d,p)$ AND $(d,d)$ REACTIONS AT 5 GeV/c AND 178 MR
\footnote{\it Talk given at the XVII-th International Baldin Seminar on
High Energy Physics Problems, 

~~ISHEPP XVII, 27 September- 2 October 2004, Dubna, Russia}}

\vskip 5mm

V.P.Ladygin$^{1,\dag}$, L.S.Azhgirey$^{1}$,
S.V.Afanasiev$^1$, V.N.Zhmyrov$^1$, L.S.Zolin$^1$, V.I.Ivanov$^1$,
A.Yu.Isupov$^1$, N.B.Ladygina$^1$,  A.G.Litvinenko$^1$, 
V.F.Peresedov$^1$, A.N.Khrenov$^1$ and N.P.Yudin$^2$ 

\vskip 5mm

{\small
(1) {\it
JINR, 141980, Dubna, Moscow region, Russia 
}
\\
(2) {\it
Moscow State University, Moscow, Russia
}
\\
$\dag$ {\it
E-mail: ladygin@sunhe.jinr.ru
}}
\end{center}

\vskip 5mm

\begin{center}
\begin{minipage}{150mm}
\centerline{\bf Abstract}
    New data on the tensor analyzing power $A_{yy}$ of the
$^9Be(d,p)X$ reaction at an initial deuteron momentum of 5 GeV/c 
and secondary particles (protons and deuterons) 
detection angle of 178 mr have been obtained at the 
JINR Synchrophasotron. 

The proton data obtained are analyzed within the 
framework of an approach based on the light-front dynamics using 
Karmanov's relativistic deuteron wave function. Contrary to the 
calculations with standard non-relativistic deuteron wave 
functions, we have managed to explain  the new data within the 
framework of our approach without invoking degrees of freedom 
additional to nucleon ones. 

The $^9Be(d,d)X$ data are obtained 
in the vicinity of the 
excitation of baryonic resonances with masses up to 
$\sim$ 1.8 GeV/$c^2$.
The  $A_{yy}$ data are
in a good agreement with the
previous data
obtained at 4.5
and 5.5 GeV/c when they are plotted versus $t$.
The results of the experiment are compared with the
predictions of the plane wave impulse approximation
and $\omega$-meson exchange models.
\end{minipage}
\end{center}

\vskip 10mm

\section{Introduction}

The interest to the $(d,p)$ reaction at relativistic energies 
is  
mostly due to the
possibility to observe
the manifestation of the non-nucleonic degrees of freedom and 
relativistic effects in the simplest bounded system.

Large amount of the polarization data in deuteron breakup obtained 
at a zero degree
last years can be interpreted from the point of view 
$NN^*$ configurations  in the deuteron, where relativistic effects
are taken into account by the minimal relativization scheme with
the dependence of the deuteron structure on single variable $k$.
In addition the considering of  multiple scattering is  required
to obtain the agreement with the data\cite{kob}.

On the other hand,
it was shown that $T_{20}$ data for the pion-free 
deuteron breakup process $dp \rightarrow ppn$ in the kinematical 
region close to that of backward elastic $dp$ scattering depended 
on the incident deuteron momentum in addition to $k$\cite{azh1}. 
The recent measurements of the tensor analyzing 
power $A_{yy}$ of deuteron inclusive breakup  on nuclear 
targets \cite{ayy90,ayy45}
have demonstrated a significant dependence on the transverse 
secondary proton momentum  $p_{T}$ being plotted at a fixed 
value of the longitudinal proton momentum.
This forces 
one to suggest that description of this quantity requires an 
additional independent variable, aside from $k$.  

At the same time, the interest to deuteron inelastic scattering off hydrogen 
and nuclei at high energies is due to the possibilities
to study nucleon-baryon ($NN^*$) interaction;
to investigate the deuteron structure at
short distances (as complementary method to the elastic
$pd$- and $ed$-scatterings, deuteron breakup reaction,
electro- and photodisintegration of the deuteron);
to learn amplitudes of $NN^*\to NN^*$ processes 
in the kinematical range, where the contribution of double scattering 
diagrams is significant;
to get information on  the formation of 6$q$ configuration in the deuteron,  
at large momentum transfers. 

Since the deuteron is an isoscalar probe,
$A(d,d')X$ reaction is selective to the isospin
of the unobserved system $X$,
which is bound to be equal to the isospin of the target $A$. 
Inelastic scattering of deuterons on hydrogen, $H(d,d')X$,
is selective to the isospin $1/2$ and
can be used to obtain  information on the
formation of baryonic resonances $N^*(1440)$,
$N^*(1520)$, $N^*(1680)$, and others.

The polarized deuterons of high energies
have been used to study  the tensor analyzing
power $T_{20}$ in the vicinity of the Roper resonance ($P_{11}(1440)$)
excitation on
hydrogen and carbon targets at Dubna
\cite{45_55} and on
hydrogen  target at Saclay \cite{morlet}.
The measurements of $T_{20}$
in the deuteron scattering at 9 GeV/c on hydrogen and carbon
have been performed for missing masses up to $M_X\sim 2.2$
GeV/c$^2$ \cite{azh9}. The experiments have shown a large negative
value of $T_{20}$ at momentum transfer of $t\sim -0.3$ (GeV/c)$^2$.
Such a behaviour of the tensor analyzing power has been interpreted
in the framework of the $\omega$-meson exchange model \cite{egle1} as due
to the longitudinal isoscalar form factor of the Roper resonance
excitation \cite{egle2}. The measurements of the tensor and vector
analyzing powers $A_{yy}$ and $A_y$ at  9 GeV/c and 85 mr of the
secondary deuterons emission angle in the vicinity of the
undetected system mass of $M_X\sim 2.2$ GeV/c$^2$ have shown
large values \cite{lad1}. The obtained results are in satisfactorily agreement with the
plane wave impulse approximation (PWIA) calculations \cite{nadia}.
It was stated that the spin-dependent part of 
the $NN\to NN*(\sim 2.2~GeV/c^2)$ 
process amplitude is significant.
The measurements of  $A_{yy}$  at 4.5 GeV/c and 80 mr \cite{lad2}
also shown large value of the tensor analyzing power. 
The exclusive measurements of the polarization observables in the
$H(d,d')X$ reaction in the vicinity of the Roper resonance excitation
performed recently at Saclay \cite{ljuda} also demonstrated large spin effects.

In this report the $A_{yy}$ data in deuteron inclusive 
breakup on beryllium at 5.0 GeV/c and 178 mr are presented~\cite{ayy50}.
The results are compared with the relativistic calculations using Paris, 
CD-Bonn and 
Karmanov's deuteron wave functions (DWFs).
Also  new results on the tensor and vector
analyzing powers $A_{yy}$ and $A_y$ in deuteron inelastic scattering
on beryllium target at the incident deuteron momentum of 5.0~GeV/c
and $\sim$178~mr of the secondary emission angle in the vicinity of 
light baryonic resonances excitation are reported \cite{ayy50dd}.

\section{Experiment}

   The experiment has been performed using a polarized deuteron
beam at Dubna Synchrophasotron at the Laboratory of High Energies
of JINR and the SPHERE setup shown in Fig.1 and described
elsewhere \cite{lad1,lad2}. The polarized deuterons were produced
by the ion source POLARIS \cite{polaris}.
The sign of the beam polarization was  changed  cyclically and
spill-by-spill, as  $"0"$, $"-"$, $"+"$, where $"0"$ means
the absence of the polarization, $"+"$ and $"-"$ correspond
to the sign of $p_{zz}$ with the quantization axis perpendicular
to the plane containing the mean beam orbit in the accelerator.

     The tensor polarization of the beam has been
determined during the experiment   
by the asymmetry of protons from the deuteron breakup on  
berillium target, $d+Be \to p+X$, at
zero emission angle and proton momentum of $p_p\sim \frac{2}{3}p_d$
\cite{zolin}. It 
was shown that deuteron
breakup reaction in such
kinematic conditions has very large tensor analyzing power
$T_{20}= -0.82\pm 0.04$, which is independent on the atomic
number of the target
($A >$ 4) and on the momentum of incident deuterons
between 2.5 and 9.0~GeV/c \cite{t20br}. The tensor polarization
  averaged over the
whole duration of the experiment   was
$p_{zz}^+=0.716\pm 0.043(stat)\pm 0.035(sys)$ and
$p_{zz}^-=-0.756\pm 0.027(stat)\pm 0.037(sys)$ in $"+"$ and $"-"$
beam spin states, respectively.

     The stability of the vector polarization of the beam has
been monitored
by measuring of the asymmetry
of quasi-elastic $pp$-scattering on thin $CH_2$ target
placed  at the $F_3$ focus of VP1 beam line. The values of the 
vector polarization were obtained using the results of the asymmetry
measurements  at the momenta  2.5~GeV/c per nucleon
and 14$^\circ$ of the proton scattering angle with
corresponding value of the effective analyzing power of the polarimeter
$A(CH_2)$  taken as $0.234$ \cite{f4}.
The vector polarization of the beam in different spin states was
$p_z^+=0.173\pm 0.008(stat)\pm 0.009(sys)$ and
$p_z^-=0.177\pm 0.008(stat)\pm 0.009(sys)$.

\begin{figure}[h]
 \epsfysize=80mm
 \centerline{
  \epsfbox{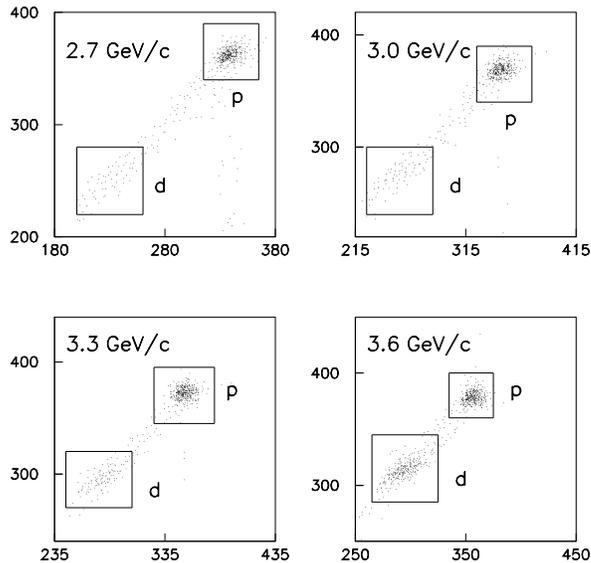}
 }
 \caption{The TOF spectra obtained for different
magnetic elements tuning. }

\end{figure}

The coincidences of signals from three
scintillation counters  were
used as a trigger.  For particle identification the time-of-flight
(TOF) information with a base line of $\sim$28 m between
the start counter and the stop counters 
 were used in the
off-line  analysis. The TOF resolution was better
than $0.2$ ns ($1\sigma$).
The TOF spectra obtained for all
four cases of magnetic elements tuning are shown in Fig.1.
At the higher momentum of the detected particles only deuterons
appear in TOF spectra, however,
when the momentum decreases the relative contribution of
protons becomes more pronounced. In data processing
useful events were selected as the ones with at least two
measured time of flight values correlated.
This allowed to
rule out  the residual background completely.

     The tensor $A_{yy}$ and vector $A_y$ analyzing powers
were calculated from
 the yields of deuterons $n^+$, $n^-$ and $n^0$ for different
states of the beam polarization after correction for dead time of the setup,
by means of the expressions
\begin{eqnarray}
\label{ayy}
A_{yy} &=& 2\cdot \frac{p_z^-\cdot (n^+/n^0-1)
~-~p_z^+\cdot (n^-/n^0-1)}{p_z^- p_{zz}^+ - p_z^+ p_{zz}^-},\nonumber\\
A_{y} &=& -\frac{2}{3}\cdot
\frac{p_{zz}^-\cdot (n^+/n^0-1)
~-~p_{zz}^+\cdot (n^-/n^0-1)}{p_z^- p_{zz}^+ - p_z^+ p_{zz}^-}.
\end{eqnarray}

\section{Results}

     The results on the tensor analyzing power $A_{yy}$
of the reaction $^9Be(d,p)X$ at the initial deuteron momentum of 5
GeV/$c$ and a proton emission angle of 178 mr are 
compared with the calculation performed in the framework of relativistic hard 
scattering model 
in Fig.~2.     The details of the calculations and
the final expressions for the tensor analyzing power of
$(d,p)$  reaction can be found 
in ref. \cite{azhyud3}.

\begin{figure}[h]
 \epsfysize=80mm
 \centerline{
  \epsfbox{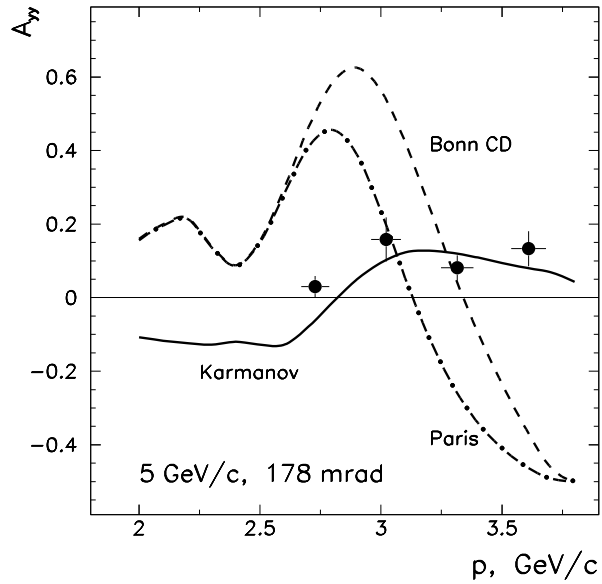}
 }
 \caption{ $A_{yy}$ in the reaction $^{9}Be(d,p)X$ at an initial
deuteron momentum of 5 GeV/$c$ and a proton emission angle of 178
mr as a function of the detected proton momentum. 
The solid curve was calculated with the Karmanov's
relativistic deuteron wave function \cite{karm1}.
The calculations
were made with the deuteron wave functions for the Bonn CD \cite{bonn}
(dashed curve) and the Paris \cite{paris} (dash-dotted curve)
potentials.  }
\end{figure}

It is seen that the
experimental data are rather good reproduced using the Karmanov's
relativistic deuteron wave function (DWF) \cite{karm1} depending on 2 internal variables
 as opposed to the calculations
with the standard DWFs \cite{paris,bonn};
the last curves change sign at the proton momentum $\sim$ 3.2 GeV/$c$.

    In Fig.3 the data on the tensor analyzing power
$A_{yy}$ in the inelastic scattering of 5.0~GeV/c deuterons on
beryllium at an angle of 178 mr are shown as a function of the
transferred 4-momentum $t$ by the solid triangles.
The $A_{yy}$ has a positive value at 
$|t|\sim 0.9$~(GeV/c)$^2$ and  crosses a 
zero at larger $|t|$. The data on tensor analyzing power
obtained at  zero emission angle
at 4.5~GeV/c and  5.5~GeV/c \cite{45_55}
on hydrogen are given by the open triangles and  squares, respectively
(recall that for these data $A_{yy} = - T_{20}/\sqrt{2}$).
The data obtained at 4.5~GeV/c and at an angle of 
80 mr \cite{lad2} are shown by the open circles.
As it was established earlier \cite{45_55,lad2}, there
is no significant dependence of $A_{yy}$ on the $A$-value of the
target. 
The observed independence of the tensor analyzing power on the
atomic number of the target indicates that the rescattering in
the target and medium effects  are small. Hence, nuclear targets
are also appropriate to obtain information  on the baryonic
excitations in 
the deuteron inelastic scattering \cite{45_55,azh9,lad1,lad2}.

\begin{figure}[h]
 \epsfysize=80mm
 \centerline{
  \epsfbox{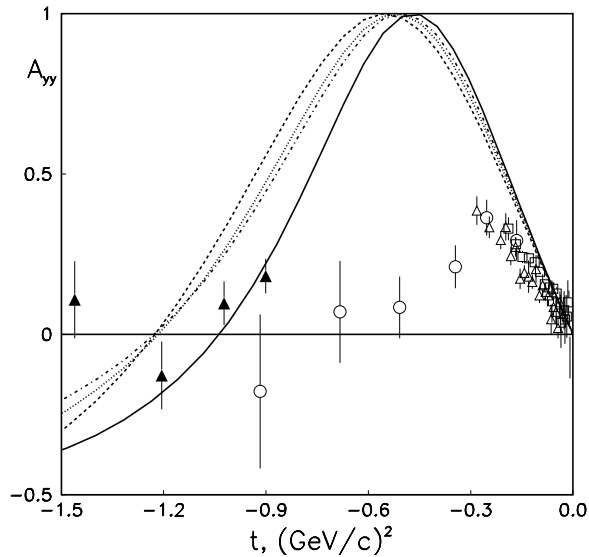}
 }
 \caption{$A_{yy}$ in  deuteron
inelastic scattering  on beryllium at 5.0 GeV/c
at an angle of 178 mr 
and at 4.5 GeV/c at an angle of 80 mr \cite{lad2} given by the full triangles and open 
circles,
respectively; 
on hydrogen at 4.5 and 5.5 GeV/c at zero angle \cite{45_55} 
shown by the open triangles and squares, respectively,
as a function of the 4-momentum $t$.
The solid, dashed, dotted
and dash-dotted lines are predictions in the framework of
PWIA \cite{nadia} using
DWFs for Paris \cite{paris} and Bonn A, B and C \cite{bonn}
potentials, respectively.
}
\end{figure}

     The sensitivity of the tensor analyzing power in the deuteron
 inelastic scattering off protons to the
excitation of baryonic resonances has been
pointed out in \cite{egle1} in the framework of the $t$-channel
$\omega$-meson  exchange model. In this model 
the $t$-dependence of the tensor analyzing power in deuteron
inelastic scattering is defined by 
the $t$-dependence of the deuteron form
factors  and the contribution of the Roper
resonance due to its nonzero isoscalar longitudinal
form factor \cite{egle2}. 
Since, the isoscalar longitudinal
amplitudes of $S_{11}(1535)$ and $D_{13}(1520)$ vanish due to spin-flavor
symmetry, while both isoscalar and isovector longitudinal couplings
of $S_{11}(1650)$  vanish identically, the tensor analyzing power $A_{yy}$ 
in inelastic deuteron scattering with the excitation 
one of these resonances has the value  of +0.25 independent of $t$ \cite{lad2}.

The $t$ dependence of $A_{yy}$ at $M_X\sim$1550~MeV/c$^2$ is 
 shown in Fig.4.
The full triangles are the results of the present experiment, 
open squares, circles
and triangles are obtained earlier at 4.5 and 5.5~ GeV/c \cite{45_55,lad2}.
The solid curves are the results of the PWIA calculations \cite{nadia} 
using Paris DWF
\cite{paris}.
The dashed lines are the expectations of the $\omega$-meson  exchange model 
\cite{egle1,egle2}.
 One can see that  the behaviour of $A_{yy}$ at $M_X\sim$ 
1550 MeV/c$^2$  is not in
contradiction with the $\omega$-meson exchange model prediction
\cite{egle2}.

\begin{figure}[hbt]
 \epsfysize=80mm
 \centerline{
  \epsfbox{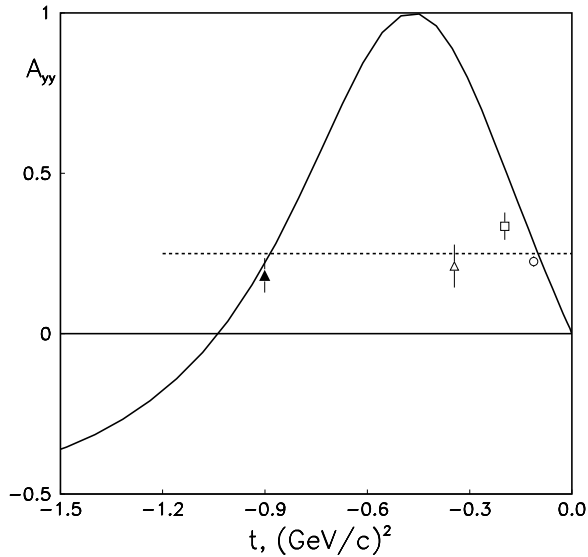}
 }
 \caption{The $A_{yy}$ data  from the present experiment (full triangles)
along with the data obtained with
4.5 and 5.5 GeV/c deuterons at zero angle \cite{45_55}
(open circles and squares, respectively) and the
data at 4.5 GeV/c at an angle of 80 mr \cite{lad2}
plotted versus 4-momentum $t$ 
 for
the missing mass $M_X\sim$1550 MeV/c$^2$.
The solid curve is the 
calculations in PWIA using
DWFs for Paris \cite{paris}.
The dashed line is the predictions within
the $\omega$-meson exchange model \cite{egle2}.
}
\end{figure}

\begin{figure}[hbt]
 \epsfysize=80mm
 \centerline{
  \epsfbox{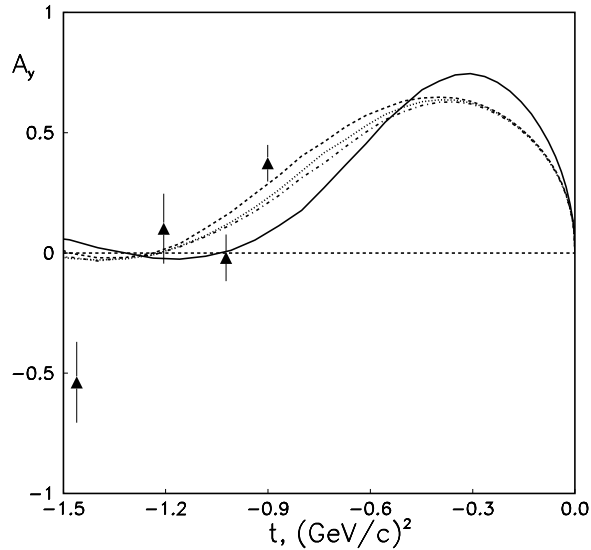}
 }
 \caption{Vector analyzing power $A_{y}$ in  deuteron
inelastic scattering  on beryllium at 5.0 GeV/c
at an angle of 178 mr 
as a function of the 4-momentum $t$.
The solid, dashed, dotted
and dash-dotted lines are predictions in the framework of
PWIA \cite{nadia} using
DWFs for Paris \cite{paris} and Bonn A, B and C \cite{bonn}
potentials, respectively. 
}
\end{figure}
The behaviour of the vector analyzing power $A_y$ obtained in 
the present experiment is plotted in Fig.5 versus $t$. The calculations
are performed 
using the expression for $A_y$ from ref.\cite{nadia} with 
the ratio $r$ of the
spin-dependent to spin-independent parts of the $NN\to NN^*$ process
taken in the form $r=a\cdot\sqrt{|t|}$ with the value of $a=1.0$.
The solid curve in Fig.5
is obtained with the DWF for Paris potential \cite{paris},
while  the dashed, dotted and dash-dotted lines
correspond to the DWFs for Bonn A, B and C potentials \cite{bonn},
respectively. 
The PWIA calculations give approximately the same results 
at the value of $a\sim 0.8\div 1.2$.  
It should be noted that $a$ value might 
have different values for the different $M_X$, however,
we took the fixed value  for the simlicity
due to lack of the data.

\section{Conclusions}

New experimental data on the tensor and vector analyzing
powers $A_{yy}$ and $A_y$
in the $^9Be(d,p)X$ and $^9Be(d,d^\prime)X$ reactions  at 5.0 GeV/c
and at an angle of $\sim$178 mr are presented.

The calculation of the tensor analyzing power of $(d,p)$ reaction
in the frame of light-front dynamics using Karmanov's relativistic
DWF is in good agreement with the new
experimental data  whereas the calculations with the standard
non-relativistic DWFs are in sharp
contradiction with the data.
New data favour the view of ref. \cite{azhyud3} that the
relation between the $k_L$ and ${\bf k}_T$ in a moving deuteron
differs essentially from that in the non-relativistic case. The
method of relativization proposed by Karmanov et al. \cite{karm1}
appear to reflect correctly this relation,
at least up to $p_T \sim$ 0.7 GeV/$c$.

     The data on $A_{yy}$ in the $(d,d^\prime)X$ reaction
are in good agreement with the data obtained in previous experiments 
at  the momenta between 4.5~GeV/c and 5.5~GeV/c \cite{45_55,lad2}
when they are compared versus variable $t$. 

It is observed also that $A_{yy}$ data in the $(d,d^\prime)X$ reaction are in good agreement
with PWIA calculations \cite{nadia} using conventional DWFs \cite{paris,bonn}.
On the other hand, 
the behaviour of the $A_{yy}$ data obtained in the vicinity of
the $S_{11}(1535)$ and $D_{13}(1520)$
resonances is  not in contradiction with the predictions of the
$\omega$-meson exchange model \cite{egle2}, while at higher excited
masses this model
may require taking into account the additional baryonic
resonances with nonzero longitudinal form factors.

The vector analyzing power $A_y$ has a large value at 
$M_X\sim$ 1500~MeV/c$^2$, that could be interpreted as a significant
role of the spin-dependent part of the elementary amplitude of the 
$NN\to NN^*$ reaction.

\begin{sloppypar}
Authors are grateful to the LHE accelerator staff and POLARIS team
for providing good conditions for the
experiment.
This work was supported in part by the Russian Foundation for Fundamental Research 
(grant No. 03-02-16224).
\end{sloppypar}

\end{document}